\documentclass[journal=jceda8,manuscript=article,layout=twocolumn]{achemso}

\usepackage[utf8]{inputenc}
\usepackage[T1]{fontenc}
\usepackage{amsmath,amssymb}
\usepackage{booktabs}
\usepackage{tabularx}
\usepackage{hyperref}

\begin{tocentry}
	\includegraphics[width=8.25 cm]{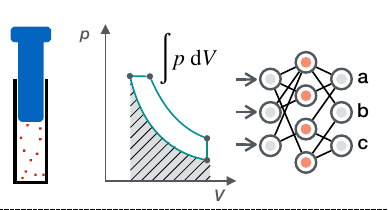}
\end{tocentry}

\author{Anna Geißler}
\affiliation[uniwue]{Institute of Physical and Theoretical Chemistry, Julius-Maximilian University W\"urzburg, 97074 W\"urzburg, Germany\vspace{0.8em}}
\author{Luca-Sophie Bien}
\affiliation[uniwue]{Institute of Physical and Theoretical Chemistry, Julius-Maximilian University W\"urzburg, 97074 W\"urzburg, Germany\vspace{0.8em}}
\author{Friedrich Schöppler}
\affiliation[uniwue]{Institute of Physical and Theoretical Chemistry, Julius-Maximilian University W\"urzburg, 97074 W\"urzburg, Germany\vspace{0.8em}}
\author{Tobias Hertel}
\affiliation[uniwue]{Institute of Physical and Theoretical Chemistry, Julius-Maximilian University W\"urzburg, 97074 W\"urzburg, Germany\vspace{0.8em}}
\altaffiliation{\small\textnormal{Benchmark downloadable at: \href{https://huggingface.co/datasets/herteltm/UTQA}{huggingface.co/datasets/herteltm/UTQA}}}
\email{tobias.hertel@uni-wuerzburg.de}
\title{From Canonical to Complex: Benchmarking LLM Capabilities in Undergraduate Thermodynamics}


\begin{document}

\begin{abstract}

Large language models (LLMs) are increasingly considered as tutoring aids in science education. Yet their readiness for unsupervised use in undergraduate instruction remains uncertain, as reliable teaching requires more than fluent recall: it demands consistent, principle-grounded reasoning. Thermodynamics, with its compact laws and subtle distinctions between state and path functions, reversibility, and entropy, provides an ideal testbed for evaluating such capabilities. Here we present UTQA, a 50-item undergraduate thermodynamics question answering benchmark, covering ideal-gas processes, reversibility, and diagram interpretation. No leading 2025-era model exceeded our 95\% competence threshold: the best LLMs achieved 82\% accuracy,  with text-only items performing better than image reasoning tasks, which often fell to chance levels. Prompt phrasing and syntactic complexity showed modest to little correlation with performance. The gap concentrates in finite-rate/irreversible scenarios and in binding visual features to thermodynamic meaning, indicating that current LLMs are not yet suitable for unsupervised tutoring in this domain.
\end{abstract}

\section{Introduction}
% motivation and focus
Large language models (LLMs) have emerged as highly capable general-purpose assistants, exhibiting impressive abilities to process, generate, and explain scientific content \cite{Hinton2006,Bengio2009,LeCun2015,Vaswani2017,Brown2020,OpenAI2023}. While much attention has focused on their potential for research support and knowledge retrieval, it remains unclear if their understanding is sufficient for unsupervised undergraduate-level tutoring. Reliable teaching requires more than fluent recall—it demands consistent reasoning that can guide students through complex problem-solving scenarios.

% why thermodynamics
Here, thermodynamics provides an excellent opportunity for deeper evaluation. Its theoretical core is compact and well-established, yet correct application demands careful separation of crucial concepts: heat vs. work, state vs. process variables, and especially reversible vs. irreversible transformations—what Duhem called “the most important and, at the same time, most problematic to be defined in Thermodynamics” \cite{Duhem1903,Hollinger1991,Bordoni2012,Callen1985,Atkins2018,Bain2014}. These distinctions are conceptually subtle yet mastering them requires both breadth and conceptual depth—spanning state functions and path dependencies.

% why a new benchmark
Despite its foundational character, existing science benchmarks devote surprisingly little attention to thermodynamic reasoning. In GPQA \cite{Rein2023}, for instance, about 80\% of chemistry questions concern organic recall, with almost no coverage of entropy or reversibility—a gap echoed in Humanity's Last Exam \cite{Phan2025}. SciBench \cite{wang2024} is closer in spirit, assembling college-level physics, mathematics, and chemistry problems, yet its thermodynamics items are limited to quantitative end-answer calculations, with little exercise of reasoning about state functions, entropy bookkeeping, or reversibility. We note that gpt-5 correctly solved all 26 SciBench items we identified as topically aligned with our focus here. Given that entropy sets the arrow of time and limits energy conversion, and that reversibility underpins the second law (called unmatched among physical principles by Einstein \cite{Einstein1949}), these omissions leave major gaps in evaluating LLM scientific reasoning.

% benchmark and threshold
We therefore introduce UTQA, a new 50-item single-choice benchmark specifically designed to evaluate LLM competence in undergraduate thermodynamics, with particular emphasis on ideal-gas processes, entropy, and reversibility. Rather than testing mere recall, our benchmark challenges models with multi-step reasoning problems including graphical representations that require integrating multiple constraints—the kind of problem-solving essential for effective tutoring. 

% initial findings
Our initial results reveal both progress and persistent limitations. Current LLMs handle many straightforward text-only items reliably, yet performance drops sharply in two critical areas: finite-rate or irreversible scenarios that require nuanced thermodynamic reasoning, and diagram-based questions that demand mapping visual features to thermodynamic meaning. Notably, no model reached our provisional reliability threshold for unsupervised instructional use of 95\% accuracy \cite{VanLehn2011}, highlighting a significant gap between fluent scientific explanation and the dependable, principle-grounded reasoning required for effective tutoring.

\section{Methods}

% Scope and interfaces.
We conducted two classes of experiments. (i) \emph{Prompting and linguistic degradation} experiments were run \emph{exclusively} on gpt-4o via the API. (ii) The \emph{cross-model benchmark comparison} was run on all other models using their command-line web interfaces under identical task settings. Full prompt texts and command invocations are provided in the Supporting Information (SI).

%Item set and modality handling.
The benchmark comprises 50 single-choice items, each with three distractors and one correct answer option—33 text-only questions and 17 diagram-based questions. Prompting and linguistic degradation experiments as well as investigations of the role of linguistic question complexity were restricted to the text-only items to isolate verbal/semantic reasoning from visual interpretation. The diagram-based questions were analyzed separately in the cross-model comparison.

% Run protocol and anti-caching precautions.
For each prompt variant, the text-only items were submitted in one or several independent runs in single-shot mode at sampling temperature $T=0.7$. To minimize context persistence and prompt-caching effects \cite{OpenAIpromptCaching}, we completed a full cycle of all 33 questions before repeating the set. This ordering suppressed spurious regularities (e.g., repetitive answer patterns) attributable to caching or internal batching rather than genuine model uncertainty.

%Prompting experiments.
We also evaluated 17 prompting strategies spanning minimal directives, suppressed-reasoning variants, structured reasoning (e.g., chain-of-thought), elimination-based prompts, persona framings, and affective/framing styles. Prompt wordings, acronyms, and attributions are listed in Table~\ref{tab:prompt-categories} and the SI. A two-phase variant (analysis followed by an explicit answer request) was also tested for comparison to single-shot prompting.
% Prompt categorization
\begin{table}[h]
\centering
\small
\begin{tabularx}{\columnwidth}{@{}l X@{}}
\toprule
\textbf{Category} & \textbf{Prompts (Acronyms)} \\
\midrule
Minimal Directive & BP, SP \\
Suppressed Reasoning & IRA, NoT \\
Reasoning Prompts & CoT, AdCoSM, Logical CoT, CoS, ToT, GoT, ThouT \\
Elimination Prompts & CoT-E, EliM \\
Persona / Expertise Amplifier & SIM \\
Affective / Framing Prompts & EI, EII, DIS \\
\bottomrule
\end{tabularx}
\caption{Prompt categories and seventeen tested prompts with associated acronyms.}
\label{tab:prompt-categories}
\end{table}

%Linguistic degradation experiments.
To quantify sensitivity to input quality, we applied controlled perturbations to a subset of items that were typically answered correctly under a simple baseline prompt. Three perturbation families were used: (i) reduced clarity (syntactic/grammatical distortions), (ii) orthographic noise (spelling/punctuation errors), and (iii) domain-terminology mismatches (systematic replacement with near-synonyms or colloquialisms). Each degraded item preserved the original physical content and correct answer key.

%Scoring and uncertainty estimation.
Accuracy per run is the fraction of correctly answered items. When multiple runs were performed, we report the mean accuracy $\bar{a}$ across runs. To separate genuine prompt effects from stochastic variability, we computed the run-to-run standard deviation $\sigma$ from residuals relative to each prompt’s mean and quote the uncertainty of the three-run mean as $\sigma_3=\sigma/\sqrt{3}$. Unless stated otherwise, comparisons between prompts are interpreted relative to $\sigma_3$.

%Reproducibility.
All question texts, diagrams, answer keys, and problem solutions, are available in the project repository (see SI).

\section{Benchmark design}

% General purpose and design of benchmarks
Benchmark datasets are central to the development and evaluation of large language models (LLMs). Internal benchmarks, curated by model developers, guide training, track progress, and ensure alignment with specific goals. Popular external benchmarks (e.g., GPQA \cite{Rein2023}, HLE \cite{Phan2025}, HolisticEval \cite{Liang2023}, LLM-SRBench \cite{Shojaee2025})—constructed independently to enable objective model comparison—have so far underrepresented thermodynamics and rarely probe core concepts such as entropy or reversibility (for a detailed breakdown of the GPQA diamond and HLE topical distributions, see SI). The SciBench benchmark \cite{wang2024} does include numerous thermodynamics items formulated as single-target calculations, but in our view it does not sufficiently probe thermodynamic reasoning, particularly around concepts such as reversibility. Our benchmark addresses these omissions directly.

% general considerations on question design
\subsection{Question Set Design}
\label{subsec:question-set-design}

\begin{figure}[htbp]
    \centering
        \includegraphics[width=8.0 cm]{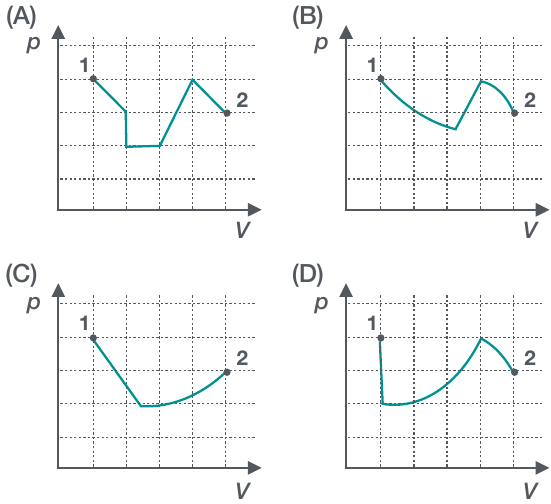}
        \caption{Representative figure from a benchmark item: four \(p\)–\(V\) diagrams depicting reversible state changes from which the case of largest pressure–volume work performed by the system must be identified.}
        \label{fig:pV-area}
\end{figure}

% design principles  
Our items were written to target clearly defined constructs in undergraduate thermodynamics while minimizing extraneous cognitive load. Each stem isolates a single concept or reasoning skill (e.g., state vs.\ path variables, sign conventions for $q$ and $w$, reversibility vs.\ finite-rate driving), and supplies only the context necessary for the intended inference. Diagrams such as the one shown in Fig. \ref{fig:pV-area} are used only when they encode essential constraints and are rendered in an instructional, low-complexity style \cite{Sweller1994}. Ambiguity is avoided by explicit units, sign conventions, and complete specifications of initial/final states or environmental contacts.

% content scope  
Content coverage spans the first and second laws, entropy changes of system and surroundings, pressure–volume work, heat transfer, and the distinction between quasistatic and non-quasistatic transformations. Items also probe recognition of path dependence, feasibility under the second law, and optimization principles in standard cycles (e.g., Carnot, Diesel). Several questions require translating between diagrammatic representations ($p$–$V$, $T$–$S$, $H$–$p$, $U$–$V$, $H$–$S$, $A$–$T$) and their thermodynamic meaning, including cases where axis labels are intentionally omitted to test structural understanding.

% formats & distractors  
Both conceptual and quantitative formats are included, from process identification to multi-step constraints that combine energy and entropy balances. Numerical problems use parameter values chosen to prevent round-off ambiguities and to ensure a unique correct option. Distractors reflect common misconceptions (e.g., equating “adiabatic’’ with “reversible’’ or confusing $dU$ with $q$), so that accuracy reflects principled reasoning rather than elimination by superficial cues.

% development & validation  
Item development followed an expert-driven, iterative process aligned with established guidance in educational measurement \cite{Haladyna2013,Messick1995}. The English set was adapted from a previously reviewed German pool and refined across multiple rounds by subject-matter experts to enforce clarity, construct focus, and unambiguous phrasing. Model solutions were independently cross-checked for correctness and pedagogical soundness. Procedural details of modality split and experimental use are described in the Methods section.

\subsection{Prompt Design and Prompting}

% scope & rationale
LLM performance can depend strongly on prompt form and linguistic quality \cite{Reynolds2021,Chatterjee2024,He2024}. We therefore varied prompt phrasing, structure, and tone to probe sensitivity on text-only items (isolating semantic reasoning from visual interpretation; see SI for protocol and uncertainty handling).

% prompt families and sources
Seventeen prompts were tested (Table~\ref{tab:prompt-categories}), spanning: minimal directives; suppressed-reasoning variants; structured reasoning prompts (CoT, ToT, GoT, Logical CoT, CoS) \cite{Wei2022,Yao2023,Besta2024,Liu2023,Hu2023}; elimination-style prompts; persona/expertise frames; and affective framings \cite{Li2023,Huang2024}. A two-phase “analyze then answer’’ format was included following \cite{Brown2020}. Full wordings and attributions are given in the SI. We included both explicit- and no-explanation variants to probe claims about unfaithful explanations and chain fragility \cite{Lanham2023,Zhang2025}, and reasoning without explicit scaffolds \cite{Kojima2022}.

% linguistic degradation manipulations
To assess robustness to input quality, we additionally applied controlled linguistic degradations to items typically solved under a baseline prompt: (i) syntactic/grammatical distortions, (ii) orthographic noise (spelling/punctuation), and (iii) domain-terminology substitutions. These edits preserved physical content and the correct key (details in Methods).

% analysis plan & references to Methods/SI
Mean accuracies were computed over independent runs with run-to-run scatter summarized by $\sigma$ and the three-run mean uncertainty $\sigma_3$ (definitions in Methods). Results by prompt family and the impact of linguistic degradation are discussed in the Results section.

\section{Results and Discussion}

\subsection{Prompting effects and uncertainty analysis}

\begin{figure}[htbp]
\centering
\includegraphics[width=8.4cm]{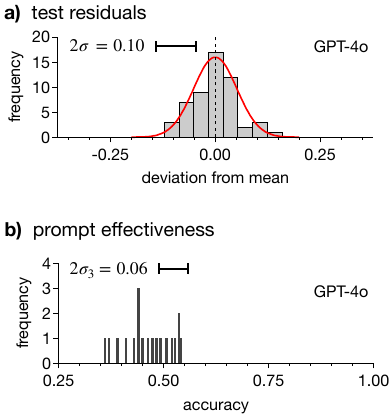}
\caption{a) Distribution of deviations from mean accuracies across 51 batch runs, corresponding to an overall spread of \(\sigma = 0.05\). b) Comparative accuracies of 17 prompting strategies; observed variation exceeds \(\sigma_3\approx0.03\).}
\label{fig:var-eff}
\end{figure}

% scope and metrics (procedure summarized; details in Methods/SI)
The 17 prompt variants were evaluated on the text-only items and quantified run-to-run scatter to separate genuine prompt effects from stochastic variability. The distribution of residuals across \(17\times3=51\) batches yields \(\sigma=0.05\), corresponding to \(\sigma_3\approx0.03\) for three-run means (Fig.~\ref{fig:var-eff}a). Mean accuracies per prompt span \(0.36\)–\(0.54\) (Fig.~\ref{fig:var-eff}b), a range that materially exceeds \(\sigma_3\), indicating real prompt sensitivity.

\begin{figure}[htbp]
    \centering
        \includegraphics[width=8.4 cm]{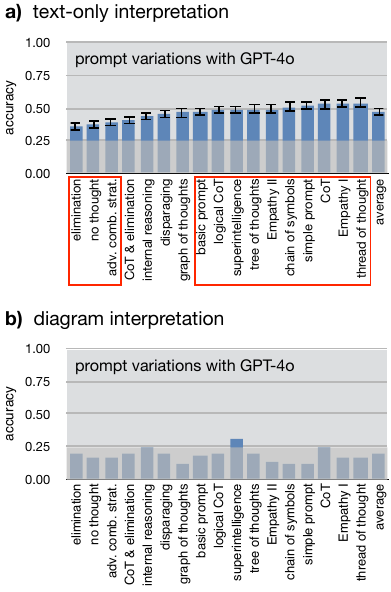}
        \caption{Accuracy scores for 17 prompting strategies using gpt-4o on 33 text-only questions. High- and low-performing groups are outlined in red. Values are three-run means; error bars indicate \(\sigma_3 \approx 0.03\).}
        \label{fig:prompting}
\end{figure}

% comparative outcomes by family (results, not mechanisms)
Sorting by accuracy (Fig.~\ref{fig:prompting}a) shows two broad bands: higher-performing variants (e.g., SP, CoT, EI, ThouT, EII, SIM) and lower-performing ones (e.g., CoT-E, NoT, AdCoSM). Minimal prompts (BP/SP) perform comparably to several structured-reasoning formats, consistent with reports that models can recruit internal reasoning without explicit scaffolds \cite{Kojima2022}. By contrast, elimination-style prompts underperform on this benchmark.

% brief explanatory discussion
Several factors likely contribute to these differences. First, longer reasoning chains introduce more opportunities for error, and mistakes that occur late in a chain may be especially damaging, a phenomenon sometimes termed “late-stage fragility” \cite{Zhang2025}. Second, explanations generated under chain-of-thought prompting may not faithfully reflect the model’s underlying computation \cite{Lanham2023}; suppressing or elaborating such reasoning (NoT, IRA, AdCoSM) therefore has limited effect. Finally, persona or affective framings can alter tone and verbosity but rarely change substantive reasoning behavior \cite{Huang2024}. Together, these points suggest that prompt design primarily affects surface presentation and stability, but does little to correct deeper deficits in scientific reasoning.

% two-phase prompting result (concise)
A two-phase “analyze then answer” format \cite{Brown2020} produced no statistically significant gains over single-shot prompting within \(\sigma_3\), suggesting that explicit separation of analysis and answer does not improve accuracy for these items using gpt-4o.

\subsection{Effect of Linguistic Degradation}

\begin{figure}[htbp]
    \centering
        \includegraphics[width=8.4 cm]{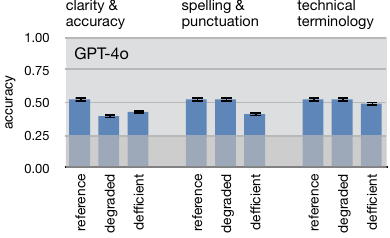}
        \caption{Accuracy for reference, degraded, and deficient versions of 33 text-only questions across three linguistic dimensions: clarity \& accuracy, spelling \& punctuation, and technical terminology. Results shown for the basic prompt; means over ten runs with error bars indicating $\sigma_{10}\!\approx\!0.015$.}
        \label{fig:linguistic-accuracy}
\end{figure}

% purpose and manipulation
We tested robustness to degraded wording under a fixed baseline prompt (\emph{Please answer the following single-choice question.}), altering only the stem and options. Controlled edits targeted three dimensions as shown in Fig. \ref{fig:linguistic-accuracy}— each at two severities (\emph{degraded}, \emph{deficient}). Edits preserved physical content and the correct answer key; the rubric used to generate variants (via gpt-4o) is provided in the SI. Table~\ref{tab:q8_degradation} illustrates all variants for a representative item.

\begin{table*}[t]
\centering
\footnotesize
\renewcommand{\arraystretch}{1.2}
\begin{tabularx}{\textwidth}{lX}
\hline
\textbf{Version} & \textbf{Wording of question} \\
\hline
Reference &
Which process of an ideal gas is characterized by the following changes:
$\Delta U = 0$, $\Delta H = 0$, and for system and environment, $\Delta G_{\text{total}} < 0$? \\
Degraded (clarity \& accuracy) &
What type of gas process shows: $\Delta U = 0$, $\Delta H = 0$, and total $\Delta G < 0$ for the system and surroundings? \\
Degraded (spelling \& punctuation) &
Which process of a ideal gas is characterised by the following changes $\Delta U = 0$, $\Delta H = 0$, and for system and environment $\Delta G_{\text{total}} < 0$ \\
Degraded (technical terminology) &
Which phase of gas behavior is marked by: $\Delta U = 0$, $\Delta H = 0$, and energy difference ($\Delta G_{\text{total}}) < 0$ for everything involved? \\
Deficient (clarity \& accuracy) &
Which kind of gas process does it when:
$\Delta U = 0$, $\Delta H = 0$, and total $\Delta G$ is going down for the gas and the rest? \\
Deficient (spelling \& punctuation) &
Wich proces of ideal gas is charactrised by these changes:
$\Delta U = 0$, $\Delta H = 0$, and for system and enviroment, $\Delta G_{\text{total}} < 0$ \\
Deficient (technical terminology) &
What kind of gas step shows:
$\Delta U = 0$, $\Delta H = 0$, and the energy change overall is negative for everything? \\
\hline
\end{tabularx}
\caption{Illustration of degradation modes for Problem 8. Variants target three factors (clarity/accuracy, spelling/punctuation, and technical terminology) at two severities (\emph{degraded} = moderate, \emph{deficient} = severe).}
\label{tab:q8_degradation}
\end{table*}

% results
Aggregate results over text-only items are shown in Fig.~\ref{fig:linguistic-accuracy}. For \emph{clarity/accuracy}, accuracy declined from $0.52$ (reference) to $0.40$ (degraded) and $0.43$ (deficient). For \emph{spelling/punctuation}, the degraded variant matched the reference ($0.52$), while the deficient variant dropped to $0.41$. For \emph{technical terminology}, the degraded score was essentially unchanged ($0.53$) and declined for the deficient variant ($0.49$). Means are based on ten runs per condition with error bars indicating $\sigma_{10}\!\approx\!0.015$.

% interpretation
Two patterns are salient for educators. First, obscuring logical structure (clarity/accuracy) leads to the largest performance losses, even at moderate severity. Second, models tolerate moderate orthographic and terminology noise, but severe errors in either dimension degrade accuracy. The weak sensitivity to terminology substitutions suggests that, for domain-familiar material, models often recover meaning from context; by contrast, degraded clarity removes essential constraints and impairs reasoning. These findings align with established comprehension factors in educational measurement \cite{Haladyna2013,Snow2009}.

\subsection{Cross-model comparison}

% setup for comparability
Using the same baseline prompt across platforms (\emph{Please answer the following single-choice question}), we evaluated 19 contemporary models (APIs and web CLI) on both the text-only and diagram-based items. 

\begin{figure}[htbp]
	\centering
		\includegraphics[width=8.4 cm]{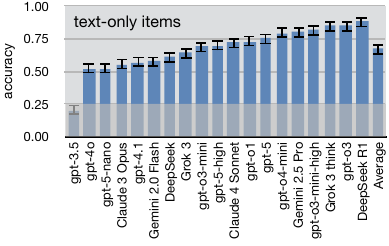}
        \caption{Comparative accuracies of different LLMs on 33 text-only interpretation questions. 
        Standard deviations of individual accuracy values are estimated at $\sigma \approx 0.05$.}
		\label{fig:rankings-text-only}
\end{figure}

% text-only results
For the text-only subset (Fig.~\ref{fig:rankings-text-only}, mean accuracies span $0.20$ (gpt-3.5-turbo-0125) to $0.88$ (DeepSeek R1), with a grand mean of $0.67$. With per-condition scatter $\sigma\!\approx\!0.05$, pairwise gaps of $\gtrsim 0.10$ are unlikely to reflect sampling variance alone. The distribution is stratified: legacy gpt-3.5 trails newer systems, while top-tier models (e.g., gpt-5 high-effort, Gemini~2.5~Pro, Grok~3~Think, gpt-o3, DeepSeek~R1) exceed $0.80$. Within families, higher “reasoning budget’’ configurations outperform lighter variants (e.g., gpt-5 high vs.\ nano), consistent with broader evidence that capacity devoted to multi-step inference improves domain problem solving \cite{Hendrycks2021b,Kojima2022}.

\subsection{Items involving diagram interpretation}

\begin{figure}[htbp]
    \centering
        \includegraphics[width=8.4 cm]{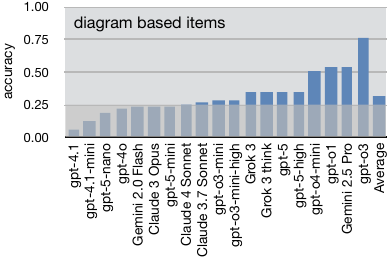}
        \caption{Comparative accuracies of all tested omni-model LLMs on the 17 diagram-based items.}
        \label{fig:rankings-diagram}
\end{figure}

Diagrams are central to thermodynamics instruction and problem solving: they externalize constraints, make path relations perceptually accessible, and support rapid inference \cite{Larkin1987,Chi1981,Schnotz2003}. Performance dropped sharply when items required interpreting such diagrams (Fig.~\ref{fig:rankings-diagram}). Across 19 models, the mean accuracy was $32\%$—less than half the text-only mean ($67\%$). The weakest result (gpt-4.1) was $6\%$ suggesting consistently chosen detractors, while the strongest (gpt-o3) reached $76\%$; Gemini~2.5~Pro and gpt-o1 scored $54\%$ and $53\%$, respectively. 

Two observations suggest that errors arise chiefly in \emph{binding} visual features to thermodynamic meaning rather than in low-level recognition. First, separate probes indicated that models typically identify axes, reference markers, and basic curve characteristics (segmentation, linearity, concavity). Second, many mistakes reflect failures to integrate diagram structure with governing relations (e.g., area under $p$–$V$ paths as work; feasibility under the second law; path ordering across segments). In contrast, trained humans can often judge work or entropy trends by quick perceptual comparison of path geometry, a classic advantage of visual reasoning \cite{Larkin1987,Chi1981}.

\subsection{Effect of linguistic complexity}

\begin{figure}[htbp]
	\centering
		\includegraphics[width=8.4 cm]{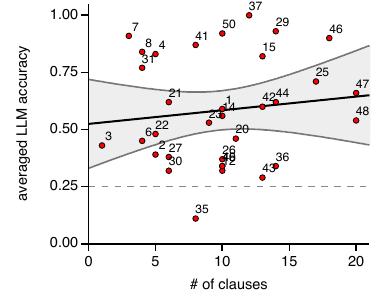}
        \caption{Average model accuracy vs.\ number of clauses for the 33 text-only items. Numerals indicate item identifiers; the dashed line marks the random-guessing baseline of 0.25. Shaded bands show $95.4\%$ confidence intervals.}
		\label{fig:clauses}
\end{figure}

Lastly we examined whether surface complexity of wording predicts accuracy on text-only items by using a simple proxy: the total number of clauses in each complete problem (stem $+$ options), where a clause contains a subject and a verb \cite{Graesser2004}. For each question we computed mean accuracy across all tested models. As shown in Fig. \ref{fig:clauses} no significant correlation emerged over the observed range of $1$–$20$ clauses.

This negative result indicates that, within typical instructional phrasing, clause count is not the limiting factor for current LLMs; failures more plausibly reflect weaknesses in conceptual integration rather than syntactic load \cite{Gibson1998}. Thus, for benchmark construction, varying clause count alone is unlikely to bias outcomes. Future analyses could evaluate richer textual metrics (e.g., referential cohesion, ambiguity) and their interaction with thermodynamic reasoning demands \cite{Graesser2004}.

\subsection{Common strengths and weaknesses}

% asymmetry across item types; single exception noted once
Across the two item classes (text-only and diagram), we observe a marked asymmetry. On text-only problems, the strongest recent models approach the level of a well-prepared graduate tutor, whereas performance on diagram-based items is broadly poor; the notable exception being gpt-o3, which consistently handles diagram-centric tasks better than other models.

% reliable strengths on text-only items
High-scoring text items share a canonical structure: they reduce to a single state-function argument or a standard identity directly applicable without multi-constraint coupling. Typical successes include recognizing that for ideal gases \(U\) and \(H\) depend only on \(T\), that throttling preserves \(H\), and that entropy increases in free expansion. Distractors encoding familiar misconceptions are routinely eliminated, indicating robust recall of first-principles relations and sign conventions.

% boundary of difficulty for text-only items (example given once)
However, accuracy degrades when solutions require integrating multiple constraints, enforcing feasibility bounds, or reasoning about finite-rate (non-quasistatic) processes. For example, in a millisecond compression of a monatomic ideal gas to \(V/2\), the bath is effectively adiabatic, and irreversible phenomena such as shock-front dissipation imply that \(T_1/T_0\) must exceed the reversible-adiabatic value \(2^{2/3}\); the strongest models increasingly identify the reversible result as a lower bound, correctly attributing the excess temperature to dissipation.

% compact list of recurring failure modes (no re-statement of example labels)
The most consistent error patterns are: (i) misuse of quasistatic templates despite explicit finite-rate cues; (ii) entropy bookkeeping errors—confusing transferred entropy with entropy production and applying \(\Delta S\) formulas outside their domain; (iii) path-dependence blind spots for work—failing to reason with oriented areas in the \(p\!-\!V\) plane and mixing sign conventions; (iv) missed invariants and feasibility constraints in optimization; and (v) numeric anchoring to textbook exponents/constants without checking applicability conditions.

% diagram interpretation: perception once, binding + composition once
On diagram-based items, many models can, when prompted, parse low-level features (axis labels, start/end states, segmentation, curvature/monotonicity). Errors arise at the \emph{binding} stage: mapping percepts to thermodynamic quantities and constraints. Recurrent issues include computing and comparing \emph{signed} areas \(\int p\,\mathrm dV\) with correct orientation, binding leg types to axes (e.g., isochoric \(\leftrightarrow\) vertical in \(p\!-\!V\)), enforcing feasibility across concatenated legs, and propagating state limits through a cycle. Template recognition (e.g., Otto/Diesel/Stirling) or simple sign-based eliminations yield acceptable performance, but tasks demanding \emph{compositional} reasoning—integrating geometry (areas, slopes, orientations) with laws and state relations—remain the principal bottleneck.

\section{Conclusions}

\begin{figure}[htbp]
	\centering
		\includegraphics[width=8.4 cm]{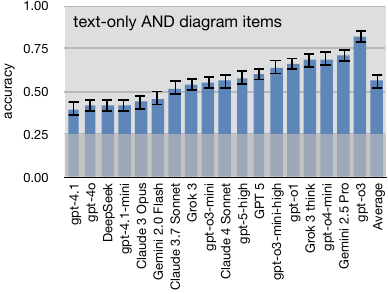}
        \caption{Omni-model comparison: overall accuracies of all tested LLMs on the complete 50-question benchmark (aggregate of text-only and diagram-related items).}
		\label{fig:rankings-all}
\end{figure}

% deceptively simple yet challenging
We introduced UTQA, a small and deceptively simple benchmark in undergraduate thermodynamics: fifty single-choice items on ideal-gas processes and thermodynamic diagrams (two thirds text-only, one third diagram-based). Applied to current LLMs, the benchmark highlights both solid performance on many canonical items and persistent weaknesses in more demanding cases. No 2025-era model reached our 95\% reliability threshold for unsupervised tutoring; even the top performer (gpt-o3 at 82\%) fell well short of this target (Fig.~\ref{fig:rankings-all}).

% dual failure modes (concise)
The performance shortfall is not uniform but concentrates in two domains. (i) Finite-rate, irreversible scenarios expose fragile regime recognition: when dissipation, feasibility bounds, or non-quasistatic driving matter, accuracies drop significantly. (ii) Diagram-based items reveal a fundamental multimodal binding deficit: models can parse axes and curve features but consistently fail to map geometry to thermodynamic meaning (e.g., signed \( \int p\,\mathrm dV \) as work, process classification, constraint consistency across cycles).

% opportunity and progress (with anthropomorphic language preserved)
These deficits are particularly striking because graphical reasoning is where humans gain considerable efficiency. However, despite current limitations, this study documents substantial progress—the strongest 2025-era systems demonstrate solid macroscopic thermodynamics with increasingly consistent microscopic narratives, representing what we would reasonably characterize as deep understanding for many items. This suggests that—contingent on improved coupling of visual perception with physical constraints—reaching our accuracy threshold on this benchmark is plausible in the foreseeable future, a promising prospect for educational applications.

% tutoring requirements + scope limits (merged)
We note that meeting accuracy thresholds is necessary but insufficient for effective tutoring. Beyond reliable problem solving, systems must also deliver appropriate interaction granularity, timely feedback, and disambiguation in dialogue—capabilities not assessed here. Our scope is intentionally narrow (ideal gases; excluding real-gas effects, mixtures, phase equilibria, and transport), so additional failure modes may emerge under broader coverage that could further challenge current capabilities.

% future directions and broader implications
Future benchmark extensions toward real-gas behavior, mixtures, phase diagrams, and standard cycles will probe reasoning under richer thermodynamic constraints and help rebalance coverage relative to other well-benchmarked domains like quantum mechanics. More generally, discipline-specific  benchmarks that encode conceptual bottlenecks provide valuable tools for measuring principled application rather than recall alone. As models improve on such carefully constructed items—and especially on multimodal binding and irreversible regimes—they move closer to becoming trustworthy, discipline-aware educational partners.

%% classes. This should be given within the "acknowledgement"

\section{Acknowledgements}

We thank Tobias Brixner, Ingo Fischer, and Roland Mitrić for their validation of a German-language predecessor to this benchmark.

\bibliography{achemso}

\end{document}